# Q-Segment: Segmenting Images In-Sensor for Vessel-Based Medical Diagnosis

Pietro Bonazzi, Yawei Li, Sizhen Bian, Michele Magno
ETH Zürich, D-ITET
Zürich, Switzerland

*Abstract*—This paper addresses the growing interest in deploying deep learning models directly in-sensor. We present "Q-Segment", a quantized real-time segmentation algorithm, and conduct a comprehensive evaluation on a low-power edge vision platform with an in-sensors processor, the Sony IMX500. One of the main goals of the model is to achieve end-to-end image segmentation for vessel-based medical diagnosis. Deployed on the IMX500 platform, Q-Segment achieves ultra-low inference time in-sensor only 0.23 ms and power consumption of only 72mW. We compare the proposed network with state-of-the-art models, both float and quantized, demonstrating that the proposed solution outperforms existing networks on various platforms in computing efficiency, e.g., by a factor of 75x compared to ERFNet. The network employs an encoder-decoder structure with skip connections, and results in a binary accuracy of 97.25 % and an Area Under the Receiver Operating Characteristic Curve (AUC) of 96.97 % on the CHASE dataset. We also present a comparison of the IMX500 processing core with the Sony Spresense, a low-power multi-core ARM Cortex-M microcontroller, and a single-core ARM Cortex-M4 showing that it can achieve in-sensor processing with end-to-end low latency (17 ms) and power consumption (254mW). This research contributes valuable insights into edge-based image segmentation, laying the foundation for efficient algorithms tailored to low-power environments.

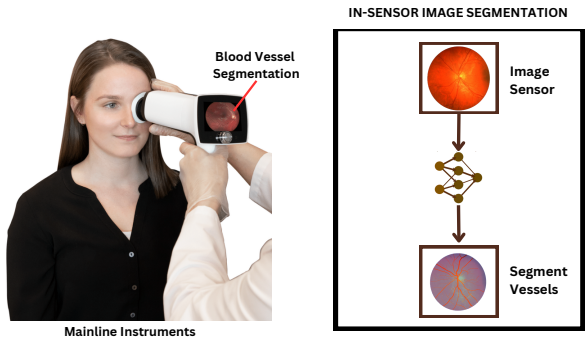

Fig. 1. End-to-end image segmentation on edge vision platforms with ultra low inference time (17 ms) and power consumption (254mW).

FUNDING & ACKNOWLEDGMENTS

This research was funded by Innosuisse (103.364 IP-ICT).

## I. INTRODUCTION

In recent years, the deployment of deep learning models on edge devices has garnered considerable attention, marking a paradigm shift in the landscape of machine learning applications [1]. This growing interest has spurred researchers to explore the efficacy of deep learning algorithms across a spectrum of single-board computers and embedded platforms [2], [3]. In particular, image processing is one of the most popular topics among researchers focusing on edge deployment. However, the convergence of image processing with edge computing introduces a myriad of challenges [4], especially given the emphasis on computational intensity and accuracy in the development of new models [5]. One primary obstacle lies in the resource constraints inherent to edge processors. In fact, these processors are characterized by limited computational memory and energy resources, rendering the deployment of resource-hungry deep learning models impractical [6]. However, recent studies [3], [7], [8] have demonstrated significant advancements in edge vision-based applications, spanning a diverse array of functionalities, ranging from the realization of autonomous drone navigation and efficient 2D gaze estimation to precise object detection on smart glasses [8].

The advent of novel edge processing technologies has ushered in a new era characterized by heightened integration, energy efficiency, and augmented computational resources. These advancements hold considerable significance as they introduce a sense of optimism amidst the challenges encountered in deploying image processing on edge devices. The ever-evolving technological landscape underscores the pivotal role of hardware components in shaping the trajectory of edge computing [9]. A notable transformative trend is the integration of cutting-edge cores directly into sensors, signifying a paradigm shift with far-reaching implications for edge processing [10], [11]. Noteworthy examples, such as the IMX500 camera from Sony [10] and the ISPU unit found in MEMS sensors by STM [11], underscore the prevalence of this trend. This integration marks a departure from convention by embedding advanced processing capabilities directly within sensors, offering the tantalizing prospect of alleviating the computational burdens traditionally borne by edge devices.

In the context of image segmentation, several lightweight segmentation algorithms have been developed. Various neural network architectures, such as MobileNetV3-Small [12],

ERFNet [13], M2U-Net [14], and T-Net [15], have been proposed to address resource constraints in edge computing scenarios. While these algorithms demonstrate effectiveness on high-end systems, there remains a notable void in evaluating their performance on low-power edge devices where the need for tiny machine learning algorithms is paramount. Our contribution lies in filling this void by conducting a comprehensive evaluation of end-to-end image segmentation algorithms on two low-power edge vision platforms. By bridging the gap in the literature, we aim to contribute valuable insights for more efficient and accessible image segmentation solutions in resource-constrained environments.

## II. RELATED WORK

### A. End-To-End Edge Vision

In response to the escalating interest in deploying deep learning models on edge devices, recent studies have diligently focused on benchmarking these algorithms across various single-board computers and embedded platforms. MCUNet [16] ran image classification at the edge with 7-10 frames per second (FPS) end-to-end. Mozhgan et al. [7] developed a vision-based autonomous drone navigation system employing the GAP8 MCU, recording a latency of 40.6 ms and an energy consumption of 34 mJ per inference. Bonazzi et al. [3] introduced, a highly efficient, fully quantized model for 2D gaze estimation, leveraging the Sony IMX500 vision sensor. Moosmann et al. [8] presented the design and implementation of a smart glasses system integrated with tiny machine learning algorithms object detection models [17], leveraging low-power processors, achieving an end-to-end latency of 56ms with 18 frames per second. Despite these advancements, a gap in the literature exists, as no comparative analysis of end-to-end image segmentation has been conducted with low-power edge platforms [18]. Unlike previous studies that primarily benchmarked image classification and specific vision-based applications on edge devices, our study delves into the intricacies of in-sensor processing and the emerging field of TinyML, contributing a nuanced understanding of how these factors influence the deployment of image segmentation models on the edge. In essence, our work proposes a segmentation algorithm in the context of low-power edge platforms, establishing a new benchmark for evaluating the efficacy of such algorithms in real-world edge computing scenarios.

### B. Lightweight Segmentation Algorithms

Various advanced neural network architectures have been proposed for image segmentation for resource constraint scenarios. MobileNetV3-Small [12], developed through a combination of hardware-aware network architecture search for low resource use cases. The ERFNet [13] introduces a deep neural network architecture for real-time (7 FPS) semantic segmentation using a Jetson TX1 [1]. M2U-Net [14] is a low-memory network, featuring a MobileNetV2 encoder part and contractive bottleneck blocks in the decoder. T-Net [15], employs dual-stream information flow, group convolutions, skip connections, and dice loss for pixel-wise classification. Despite this advancement, a notable gap remains stemming from the lack of evaluation of algorithms on low-power edge devices which truly need tiny machine learning algorithms. While the mentioned architectures demonstrate their effectiveness on high-end systems and powerful embedded devices like Jetson TX1, our work introduces a comprehensive evaluation of two low-power edge vision platforms. With the proposed Q-Segment, we achieve, for the first time, in-sensor segmentation by combining quantization with a novel processing unit directly embedded in the camera sensors [3], [10].

## III. METHODOLOGY

### A. Network Architecture

Our architecture aims to capture hierarchical features for accurate segmentation, similar to [15]. The network architecture follows an encoder-decoder structure with skip connections and specular hidden dimensions (16, 32, 64). Every block (green in Fig .2, "ConvBlock" in Table I) consists of a 3x3 convolutional layer with batch normalization and ReLu activation, followed by a 1x1 convolutional and a depth-wise convolution block. Depending on the type of the block (encoding/decoding) max pooling/unpooling operations are applied. Skip connections are placed both between corresponding encoder and decoder modules and before and after each block in order to facilitate the information flow. The output segmentation map is produced by a final convolutional layer. The selection of convolution kernels is intentional, guided not only by the inherent effectiveness of convolutional neural networks in image-related tasks but also by the demonstrated efficiency of in-sensor processing within the IMX500, especially with integer operations. It is noteworthy that the process of quantization, a crucial aspect in this context, does not adversely impact the performance of the CNN, further emphasizing the robustness and efficacy of our chosen approach.

TABLE I
THE NETWORK CONFIGURATION FOR EACH LAYER.

| Layer | Location | Type | $c_{in}$ | $c_{out}$ |
|---|---|---|---|---|
| 1 |  | ConvBlock + Pool | 3 | 16 |
| 2 | Encoder | ConvBlock + Pool | 16 | 32 |
| 3 |  | ConvBlock + Pool | 32 | 64 |
| 4 | Intermediate | ConvBlock | 64 | 64 |
| 5 |  | ConvBlock + UnPool | 64 | 32 |
| 6 | Decoder | ConvBlock + UnPool | 32 | 16 |
| 7 |  | ConvBlock + UnPool | 16 | 16 |
| 8 |  | Conv | 16 | 1 |

### B. Data Preparation

We trained our model on the CHASE [19] dataset, which was divided into training and validation sets with a ratio of 20:8, respectively. Given the small size of CHASE, data augmentation is crucial for training deep neural networks to generalize well. Various augmentation techniques were applied during training, including brightness adjustments (with a factor of 0.2), random horizontal flips (with a probability of 0.5), random vertical flips (0.5), and random rotations within the range of -1 to 1 degree. The pixel intensities of the images were normalized to the range [0, 1] to ensure numerical stability during training.

---
[1] https://developer.nvidia.com/embedded/jetson-tx1

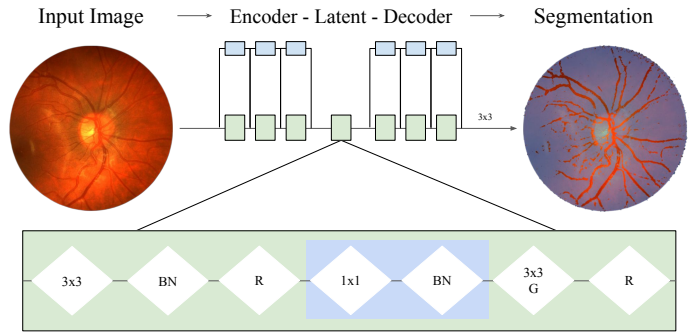

Fig. 2. A detailed diagram of the architecture and the ConvBlock (green).

*C. Loss Function*

The objective function of our network is inspired by PraNet [20] and aims to measure the dissimilarity between the predicted segmentation and the ground truth. It is defined as a combination of weighted binary cross-entropy (WBCE) and weighted intersection over union (WIOU). The weight calculation involves the absolute difference between the average-pooled ground truth and the original ground truth. The final loss is a mean value of the weighted binary cross-entropy and the weighted intersection over union. This objective function is designed to guide the training process toward accurate and spatially meaningful segmentation results.

*D. Training Configuration*

The training process involved a batch size of 8 samples and conducting 4000 epochs. The model was trained with automatic mixed precision, which automatically employs lower-precision data types when possible without sacrificing model accuracy. This not only optimized memory usage but also expedited training times.

We utilized stochastic gradient descent (SGD) for gradient optimization, setting the initial learning rate at 1e-3. To further refine the training dynamics and prevent convergence to sub-optimal solutions, a scheduler incorporating cosine annealing with warm restarts was employed. The warm restarts occurred every 20 epochs, enhancing the model's adaptability to varying complexities within the training data. This combination of techniques aimed to strike a balance between efficient convergence and robust generalization.

## IV. INFERENCE HARDWARE

In this paper, we present a comparative study of two state-of-the-art platforms for computer vision applications: the Sony IMX500 and the Sony Spresense. The Sony IMX500 [2] is a novel image sensor with a stacked pixel architecture and an integrated AI processor, which eliminates the need for external memory. The IMX500 is capable of performing real-time, in-sensor processing, making it ideal for AI applications that require low latency and high throughput. The Sony Spresense [3], on the other hand, is a versatile platform that features a powerful ARM Cortex-M4F CPU, 6 cores, and a dedicated camera interfac. The Spresense is designed to be highly flexible and can be used in a wide range of edge computing scenarios. In this paper, we evaluate the performance of these two platforms on medical imagining segmentation.

## V. EXPERIMENTAL RESULTS

The performance of our proposed network is compared with other implementations in Table II. The table shows that our network has a lower number of parameters than MN-V3-S, ERFNet, and M2U-Net. Our network also has a smaller size than all other networks except T-Net. These results demonstrate that our network is extremely efficient and compact compared to other networks.

TABLE II
COMPARISON RESULTS WITH OTHER IMPLEMENTATIONS

| Network | Params (M) ↓ | Size (Mb) ↓ |
|---|---|---|
| MN-V3-S | 2.5 | 11.0 |
| ERFNet | 2.06 | 8.0 |
| M2U-Net | 0.55 | 2.2 |
| T-Net | **0.026** | 0.11 |
| Q-Segment | 0.104 | **0.10** |

[2] https://developer.sony.com/imx500/
[3] https://developer.sony.com/spresense/

## A. System Performance

Table III and Figure 3 provides a comparative overview of the performance metrics for several state-of-the-art embedded semantic segmentation implementations. Notably, the Q-Segment implementation on the Sony IMX-500 platform stands out with an impressive latency of 0.23 ms and an exceptionally low power consumption of 72 mW. This combination of ultra-low latency and minimal power usage positions the Sony IMX-500 as a promising platform for real-time semantic segmentation applications.

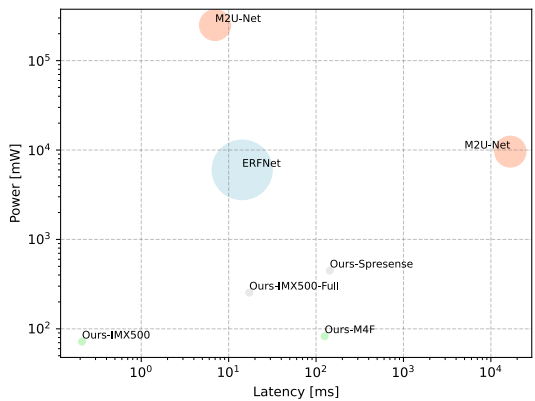

Fig. 3. A scatter chart of the embedded segmentation implementations. The size of the circles is directly proportional to the parameters size.

TABLE III
A COMPARISONS OF THE EMBEDDED SEGMENTATION IMPLEMENTATIONS.

| Method | Platform | L [ms] ↓ | P [mW] ↓ |
|---|---|---|---|
| ERFNet [13] | NVIDIA Jetson | 14.28 | 6'000 |
| M2U-Net [14] | Rockchip RK3399 | 14700 | 9'600 |
| T-Net [15] | NVIDIA GTX 1080T | N/A | 250'000 |
| Q-Segment | ARM Cortex-M4F | 126.25 | 82.69 |
|  | Sony Spresense | 143.75 | 448.00 |
|  | IMX-500 | **0.23** | **71.94** |
|  | IMX-500-Full | 17.21 | 254.60 |

## B. Model Performance

In the realm of lightweight networks for medical image segmentation, T-Net [15] stands out as a noteworthy model, showcasing superior performance on the CHASE dataset in terms of Dice coefficient, accuracy (Acc), and area under the curve (AUC). T-Net outperforms other recent lightweight networks such as M2U-Net, ERFNet, and MN-V3-S in various metrics, as illustrated in Table IV.

Our model borrows the same architecture of T-Nets with reduce memory footprint thanks to quantization. Q-Segment surpasses existing lightweight networks alternatives and is compatible with low power edge platforms.

TABLE IV
COMPARISON RESULTS W.R.T RECENT LIGHT-WEIGHT NETWORKS IN TERMS OF QUANTITATIVE PERFORMANCE.

| Dataset | CHASE | | | |
|---|---|---|---|---|
| Method | Resolution | Dice ↓ | Acc ↓ | AUC ↓ |
| M2U-Net [14] | 960x960 | 0.8006 | 0.9703 | 0.9666 |
| ERFNet [13] |  | 0.7872 | 0.9716 | 0.9785 |
| MN-V3-S [12] |  | 0.6837 | 0.9571 | 0.9673 |
| T-Net [15] |  | 0.8143 | 0.9739 | 0.9889 |
| Q-Segment |  | 0.8023 | 0.9725 | 0.9697 |

The table in Figure V showcases the performance of our retinal vessel segmentation model on images from the validation set. The qualitative results highlight the accuracy of predictions compared to ground truth across different inputs.

TABLE V
QUALITATIVE RESULTS ON IMAGES FROM THE VALIDATION SET

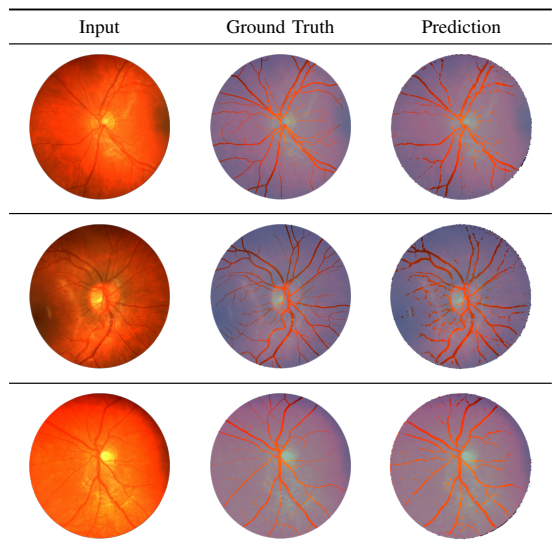

| Input | Ground Truth | Prediction |

## VI. CONCLUSION

In summary, this paper proposed Q-Segment, a T-Net inspired model which outperforms existing networks in image segmentation tasks, and exhibits superior efficiency in terms of size and parameter count. Deployed on the energy-efficient IMX500 platform, the model demonstrates a compelling reduction in power consumption to 73mW compared to other system implementations using NVIDIA GTX 1080T (250W) and a remarkable decrease in latency (530FPS vs 7FPS in ERFNet). The system performance analysis underscores Q-Segment's effectiveness on resource-constrained edge devices, particularly the IMX-500 platform, with significantly lower end-to-end latency (17 ms) and end-to-end power consumption (254 mW) compared to alternative low-power edge vision platforms (Spresense).